\documentstyle[12pt,fleqn]{article}

\def\op#1{\mathop{{\it\fam0} #1}\limits}

\newcommand{\df}{{\rm def}}

\newcommand{\beq}{\begin{equation}}
\newcommand{\eeq}{\end{equation}}
\newcommand{\ben}{\begin{eqnarray}}
\newcommand{\een}{\end{eqnarray}}
\newcommand{\be}{\begin{eqnarray*}}
\newcommand{\ee}{\end{eqnarray*}}
\newcommand{\bea}{\begin{eqalph}}
\newcommand{\eea}{\end{eqalph}}

\newcounter{eqalph}
\newcounter{equationa}
\newcounter{example}
\newcounter{remark}
\newcounter{theorem}
\newcounter{proposition}
\newcounter{lemma}
\newcounter{corollary}
\newcounter{definition}

\def\thedefinition{\arabic{definition}}

\newenvironment{theo}{\refstepcounter{definition} \medskip\noindent{\bf
Theorem \thedefinition.}\sl}{\medskip }
\newenvironment{prop}{\refstepcounter{definition} \medskip\noindent{\bf
Proposition \thedefinition.}\sl}{\medskip }

\newenvironment{eqalph}{\stepcounter{equation}
\setcounter{equationa}{\value{equation}}
\setcounter{equation}{0}

\begin{eqnarray}}{\end{eqnarray}\setcounter{equation}{\value{equationa}}}

\begin{document}

\begin{center}

{\large \bf Remark on the Serre--Swan theorem for non-compact manifolds}
\bigskip

{\sc G.Sardanashvily}
\medskip

{\small Department of Theoretical Physics, Moscow State University, 117234
Moscow, Russia

E-mail: sard@grav.phys.msu.su}

\end{center}

\bigskip

{\small
\centerline{\bf Abstract}
\medskip

The Serre--Swan theorem provides the link between projective modules of
finite rank and vector bundles over compact 
manifolds, and plays a prominent role in non-commutative geometry. Its
extension to non-compact manifolds is discussed.
}

\bigskip
\bigskip

The Serre--Swan theorem had been established for continuous functions
and sections \cite{swan}, but then it was extended to the smooth case
(see Proposition 4.1 in Ref. \cite{land} and Proposition 3.1,
Theorem 3.2 in Ref. \cite{var}). 

Recall that smooth manifolds are assumed to be real,
finite-dimensional, Hausdorff,  
second-countable, and connected. Vector bundles are finite-dimensional.

\begin{theo} \label{sp60} (Serre--Swan). 
(i) The $C^\infty(X)$-module of global sections of a smooth
vector bundle
$E\to X$ over a compact manifold $X$ is a projective module of finite rank.

(ii) Any projective $C^\infty(X)$-module of finite rank is isomorphic
to the module of global sections of some smooth vector bundle over $X$.
\end{theo}

In non-commutative geometry, one therefore thinks of a finite
projective $*$-module over a dense unital $*$-subalgebra of a
$C^*$-algebra as being a non-commutative vector bundle.

Note that item (i) of Theorem \ref{sp60} is equivalent to the
following well-known theorem (see Theorem 6.5 in Ref. 
\cite{karo} in the continuous case).

\begin{theo} \label{ss1} 
Let $E$ be a vector bundle over a compact manifold $X$. There exists a
vector bundle $E'\to X$ such that the Whitney sum $E\oplus E'$ is a
trivial vector bundle.
\end{theo}

The key point of the proof of item (i) of Theorem \ref{sp60} is that any
vector bundle 
over a compact manifold $X$ admits a bundle atlas over a finite covering
of $X$. Proposition IX in Ref. \cite{greub} generalizes this assertion
to non-compact manifolds as follows.

\begin{prop} \label{sp2} 
A smooth fibre bundle $Y\to X$ over an arbitrary manifold $X$ 
admits a bundle atlas over a finite
covering of $X$.
\end{prop}

Its proof is based on the fact that, 
for any covering of a manifold $X$, there exists a refinement
$\{U_{ij}\}$, where $j$ and $i$ run through a countable set and a
finite set, respectively, such that 
$U_{ij}\cap U_{ik}=\emptyset$, $j\neq k$. 
Let $\{U_\xi,\psi_\xi\}$ be a bundle atlas of a fibre bundle $Y\to X$ over a
covering $\{U_\xi\}$ of $X$. Let $\{U_{ij}\}$ be the above mentioned
refinement of this covering and 
$\{(U_{ij}, \psi_{ij})\}$ the corresponding bundle atlas of 
$Y\to X$. Then $Y\to X$ has the finite bundle atlas 
\beq
U_i\op=^\df\op\cup_j U_{ij}, \qquad \psi_i(x)\op=^\df\psi_{ij}(x), \qquad x\in
U_{ij}\subset U_i. \label{ss2}
\eeq

It is readily observed that, if $Y\to X$ is a vector bundle,
the atlas (\ref{ss2}) is an atlas of a vector bundle. It follows that every
smooth vector bundle admits a finite atlas, and
the proof in Refs. \cite{land,var} of item (i) of Theorem
\ref{sp60} can be generalized straightforwardly to
non-compact manifolds. Similarly, the above
mentioned Theorem 6.5 in \cite{karo} can be extended to non-compact
finite-dimensional topological manifolds. The proof of item (ii) in Theorem
\ref{sp60} does not imply the compactness of $X$.


\begin{thebibliography}{ederf}


\bibitem{greub} W.Greub, S.Halperin and R.Vanstone, {\it Connections,
Curvature, and Cohomology}, Vol. 1,  (Academic Press, N.Y., 1972).

\bibitem{karo} M.Karoubi, {\it $K$-Theory. An Introduction} (Springer-Verlag,
Berlin, 1978).

\bibitem{land} G.Landi, An introduction to non-commutative spaces and
their geometry, E-print arXiv: hep-th/9701078.

\bibitem{swan} R.Swan, Vector bundles and projective modules, {\it
Trans. Amer. Math. Soc.} {\bf 105} (1962) 264.

\bibitem{var} J.V\'arilly and J.Grasia-Bondia, Connes' noncommutative
differential geometry and the standard model, {\it J. Geom. Phys.} {\bf
12} (1993) 223.

\end{thebibliography}
\end{document}